\documentclass[aps,prl,twocolumn,showpacs,groupedaddress,floatfix]{revtex4-1}

\usepackage{graphicx}
\usepackage{dcolumn}
\usepackage{bm}
\usepackage{amsmath}
\usepackage{amssymb}
\usepackage{url}
\usepackage[normalem]{ulem}
\usepackage{makecell}
\usepackage{multirow}

\newcommand{\beq}{\begin{eqnarray}}
\newcommand{\eeq}{\end{eqnarray}}

\newcommand{\real}{{\sf I}\kern-.12em{\sf R}}
\newcommand{\comp}{{\sf I}\kern-.50em{\sf C}}
\newcommand{\unity}{{\sf I}\kern-.54em{\sf 1}}

\def\spose#1{\hbox to 0pt{#1\hss}}
\def\ltapprox{\mathrel{\spose{\lower 3pt\hbox{$\mathchar"218$}}
 \raise 2.0pt\hbox{$\mathchar"13C$}}}

\begin{document}

\title{Critical line of 2+1 flavor QCD: Toward the continuum limit}
\author{Paolo Cea}
\affiliation{Dipartimento di Fisica dell'Universit\`a di Bari 
and INFN - Sezione di Bari, I-70126 Bari, Italy}
\email{paolo.cea@ba.infn.it}
\author{Leonardo Cosmai}
\affiliation{INFN - Sezione di Bari, I-70126 Bari, Italy}
\email{leonardo.cosmai@ba.infn.it}
\author{Alessandro Papa}
\affiliation{Dipartimento di Fisica dell'Universit\`a della Calabria \\
and INFN - Gruppo collegato di Cosenza, I-87036 Arcavacata di Rende, 
Cosenza, Italy}
\email{papa@cs.infn.it}

\date{\today}

\begin{abstract}
We determine the continuum limit of the curvature of the pseudocritical line 
of QCD with $n_f$=2+1 staggered fermions at nonzero temperature and quark 
density. We perform Monte Carlo simulations at imaginary baryon chemical
potentials, adopting the HISQ/tree action discretization, as implemented in 
the code by the MILC collaboration. Couplings are adjusted so as to move on 
a line of constant physics, as determined in Ref.~\cite{Bazavov:2011nk}, with 
the strange quark mass $m_s$ fixed at its physical value and a light-to-strange 
mass ratio $m_l/m_s=1/20$. The chemical potential is set at the
same value for the three quark species, $\mu_l=\mu_s\equiv \mu$.
We attempt an extrapolation to the continuum using the results 
on lattices with temporal size up to $L_t=12$. Our estimate for the
continuum value of the curvature $\kappa$ at zero baryon density, 
$\kappa=0.020(4)$, is compared with recent lattice results and with 
experimental determinations of the freeze-out curve. 
\end{abstract}

\pacs{11.15.Ha, 12.38.Gc, 12.38.Aw}

\maketitle

\section{Introduction}
\label{introd}

The phase diagram of strongly interacting matter in the temperature ($T$) - 
baryon density plane remains a challenge for theoretical physics. Although there
is little doubt that it features a low-temperature hadronic phase, with 
broken chiral symmetry, and a high-temperature deconfined phase,
with restored chiral symmetry, the question about the precise location
and the exact nature of the transition between these two phases is still open. 
Yet, the answer to this question has many phenomenological implications:
the region of the phase diagram with high $T$ and small baryon density is
relevant for the physics of the early Universe, whereas the region of 
low $T$ and high baryon density is interesting for the astrophysics of 
some compact objects, but other corners of the phase diagram are not less 
interesting (see Ref.~\cite{Stephanov:2007fk} for an overview). 

Relativistic heavy-ion collisions provide us with a unique opportunity
to infer properties of the transition: depending on the energy of the ion beams
and on the mass number of the colliding ions, the fireball generated in the 
collision could fulfill the temperature and baryon density conditions under
which the deconfined phase appears as a transient state, before the
system freezes out into hadrons, which are then detected. Thermal-statistical 
models, assuming approximate chemical equilibrium at the chemical freeze-out
point, are able to describe the particle yields at a given collision energy
in terms of two parameters only, the freeze-out temperature $T$ and the baryon 
chemical potential $\mu_B$. The collection of freeze-out parameters extracted
from experiments with different collision energy lie on a curve in 
the $(T,\mu_B)$-plane, extending up to $\mu_B\lesssim$ 800 MeV (see Fig.~1 of 
Ref.~\cite{Cleymans:2005xv}, or Ref.~\cite{Becattini:2012xb} for a recent 
re-analysis of experimental data). 

Quantum ChromoDynamics (QCD) is widely accepted as the theory of strong 
interactions and, as such, must encode all the information needed to
precisely draw the phase diagram in the $(T,\mu_B)$-plane. As a matter of
fact, only some corners of it can be accessed by first-principle applications
of QCD, in the perturbative or in the nonperturbative regime. Here we 
focus on the lattice approach of QCD, based on the idea of discretizing
the theory on a Euclidean space-time lattice and simulating it by Monte Carlo 
numerical simulations as a statistical system, with Boltzmann weight 
given by $\exp(-S_E)$, where $S_E$ is the QCD Euclidean action. The region
of the phase diagram where $\mu_B/(3 T)\lesssim 1$ is within the reach of this 
approach and one can therefore address, at least inside this region, the 
problem of determining the shape taken by the QCD pseudocritical line 
separating the hadronic from the deconfined phase.

There is no {\it a priori} argument for the coincidence of the QCD 
pseudocritical line with the chemical freeze-out curve: if the deconfined 
phase is realized in the fireball, in cooling down the system first 
re-hadronizes, then reaches the chemical freeze-out. This implies that the 
freeze-out curve lies below the pseudocritical line in the $\mu_B$-$T$ plane. 
It is a common working hypothesis that the delay between chemical freeze-out and 
rehadronization is so short that the two curves lie close to each other and 
can therefore be compared.
Under the assumptions of charge-conjugation invariance at $\mu_B=0$ and 
analyticity around this point, the QCD pseudocritical line, as 
well as the freeze-out curve, can be parameterized, at low baryon densities, 
by a lowest-order  expansion in the dimensionless  quantity $\mu_B/T(\mu_B)$, as
\begin{equation}
\frac{T(\mu_B)}{T_c(0)}=1-\kappa \left(\frac{\mu_B}{T(\mu_B)}\right)^2 \, + \, \ldots \;,
\label{curv}
\end{equation}
where $T_c(0)$ and $\kappa$ are, respectively, the pseudocritical temperature 
and the curvature at vanishing baryon density.

Direct Monte Carlo simulations of lattice QCD at nonzero baryon density
are hindered by the well known ``sign problem'': $S_E$ becomes complex and 
the Boltzmann weight loses its sense. Several ways out of this problem
have been devised (see Ref.~\cite{Philipsen:2005mj,*Schmidt:2006us,*deForcrand:2010ys,*Aarts:2013bla} for a review): 
redesigning the Monte Carlo updating algorithms for a complex action~\cite{Karsch:1985cb,*Aarts:2009uq,*Aarts:2011ax,*Aarts:2011zn,*Seiler:2012wz,*Aarts:2013uxa,*Sexty:2013ica}, 
reweighting from the ensemble at $\mu_B=0$~\cite{Barbour:1997bh,*Fodor:2001au,*Fodor:2001pe,*Fodor:2004nz}, 
Taylor expanding the relevant observables around $\mu_B=0$ and calculating the 
first coefficients of the series by simulations at $\mu_B=0$~\cite{Gottlieb:1988cq,*Choe:2001cq,*Choe:2002mt,*Allton:2002zi,Allton:2003vx,*Allton:2005gk,*Ejiri:2005uv,Endrodi:2011gv,*Borsanyi:2012cr}, 
using the canonical formulation~\cite{Alford:1998sd,*Hasenfratz:1991ax,*Kratochvila:2005mk,*Alexandru:2005ix,deForcrand:2006ec},
using the density of states method~\cite{Bhanot:1986kv,*Karliner:1987cu,*Azcoiti:1990ng,*Ambjorn:2002pz}
and simulating the theory at imaginary chemical potentials and performing
the analytic continuation to real ones~\cite{Lombardo:1999cz,*deForcrand:2002ci,*deForcrand:2003hx,*D'Elia:2002gd,*D'Elia:2004at,*Azcoiti:2005tv,*deForcrand:2006pv,*deForcrand:2008vr,*deForcrand:2007rq,Cea:2010md,Cea:2010fh,*Cea:2012vi,Cea:2012ev,*Wu:2006su,*Nagata:2011yf,*Giudice:2004se,*Giudice:2004pe,*Papa:2006jv,*Cea:2007wa,*Cea:2007wa,*Cea:2010bp,*Cea:2009ba,*Cea:2010bz,*Karbstein:2006er,Cea:2014xva,Bonati:2014rfa,Bonati:2015bha,Bellwied:2015rza}.

The numerical evidence gathered so far in QCD with $n_f=2+1$ and physical 
or almost physical quark masses points to a scenario with a smooth
crossover between the hadronic and the deconfined (or chirally symmetric)
phase at $\mu_B=0$, with a pseudocritical temperature $T_c(0)$ of about 
155~MeV~\cite{Aoki:2006br,*Aoki:2006we,*Aoki:2009sc,Bazavov:2011nk,*Bhattacharya:2014ara}. This crossover behavior should persist in some neighborhood of
$\mu_B=0$, up to the onset of a first-order transition at some value of 
$\mu_B>0$. 

The state-of-the-art of lattice determinations of the curvature $\kappa$,
up to the very recent papers of Ref.~\cite{Bonati:2015bha,Bellwied:2015rza}, 
is summarized in Fig.~10 of Ref.~\cite{Bonati:2014rfa}: depending on the
lattice setup and on the observable used to probe the transition, the value 
of $\kappa$ can change even by almost a factor of three. The lattice 
setup dependence stems from the kind of adopted discretization, the lattice 
size, the choice of quark masses and chemical potentials, the procedure
to circumvent the sign problem. This dependence would totally disappear if, 
ideally, all groups would use the same lattice setup. A contribution to the
understanding of the impact of the lattice setup dependence is provided 
in the Appendix~B of Ref.~\cite{Bonati:2014rfa}. The dependence on the
probe observable is, instead, irreducible: since a smooth crossover is taking
place rather than a true phase transition, one cannot define a {\it bona fide} 
order parameter whose behavior would permit to uniquely locate the transition
point; instead, for any adopted surrogate observable, a different transition
point should be expected, at least in principle.

On the side of the determinations of the freeze-out curve, two recent 
determinations~\cite{Cleymans:2005xv,Becattini:2012xb} of $\kappa$, both
based on the thermal-statistical model, but the latter of them including the
effect of inelastic collisions after freeze-out, give two quite different
values of $\kappa$, each seeming to prefer a different subset of lattice
results (see Fig.~3 of Ref.~\cite{Cea:2014xva} for a snapshot of the
situation). 

The aim of this work is to contribute to a better understanding of the 
systematics underlying lattice determinations of the curvature $\kappa$, by 
corroborating our previous determination~\cite{Cea:2014xva} with 
an extrapolation to the continuum limit and by comparing it with
experimental analyses of the freeze-out curve.

Our lattice setup is as follows.  We simulate the HISQ/tree action of the 
MILC collaboration with 2+1 staggered fermions on lattices with temporal
extension $L_t=6$, 8, 10 and 12 and aspect ratio equal to four. We work on 
the line of constant physics (LCP) as determined in Ref.~\cite{Bazavov:2011nk},
with the strange mass set at the physical value and the light quark mass fixed 
at $m_l=m_s/20$. 
As discussed in Ref.~\cite{Bazavov:2011nk}, this amounts to tune the strange quark mass 
until the mass of the fictitious $\eta_{s\bar{s}}$ meson matches the lowest order perturbation theory estimate
$m_{\eta_{s\bar{s}}} = \sqrt{2 m_K^2 - m_\pi^2}$. Consequently within our simulations the pion mass is 
$m_\pi \simeq 160 \, \text{MeV}$.

We perform simulations at imaginary quark chemical potentials, 
assigning the same value to the three quark species, $\mu_l=\mu_s\equiv \mu$, 
then extrapolate to real chemical potentials. Our probe observables are the 
disconnected susceptibility of the light quark chiral condensate and its 
renormalized counterpart.
Simulating the theory at imaginary chemical potentials poses no restriction
on the lattice size or in the choice of the couplings. However, the periodicity 
in ${\rm Im}(\mu_l)/T$ of the partition function~\cite{Roberge:1986mm} implies 
that the information gathered outside a narrow interval of imaginary chemical
potentials is redundant. For the setup with $\mu_l=\mu_s\equiv \mu$, this 
interval can be chosen as the region $0\leq {\rm Im}(\mu)/T\leq \pi/3$. A safe
extrapolation of the critical line to real chemical potentials requires that
it exhibits a smooth dependence on imaginary chemical potential over this
interval, a condition which must be checked to be satisfied {\it a posteriori} 
by our data. 

The preference to the disconnected susceptibility of the light quark chiral 
condensate has  multiple motivations~\cite{Bazavov:2011nk}.  First of all, for 
small enough quark masses, its contribution to the chiral susceptibility 
dominates over the connected one, which is harder to compute; then, it shows a 
strong sensitivity to the transition; finally, it is exempt from additive 
renormalization, undergoing only a multiplicative one. This translate into
a very precise determination of the critical couplings at imaginary $\mu$,
which is the main prerequisite of a safe extrapolation to the real values of
$\mu$.

There are two main limitations in our setup. The first is that we work
with a physical strange quark mass, but with light quarks a bit heavier
than physical ones. Numerical results in $n_f=2$ indicate a mild dependence 
of the curvature on the quark mass (see the discussion in Sec.~III of 
Ref.~\cite{Cea:2012ev}). If the same applies here, as we believe, our result
for $\kappa$ will only slightly underestimate (in absolute value) the true 
physical curvature. The second limitation is that, for the sake of comparison
with the freeze-out curve, our setup of chemical potentials could not be the
one which better reproduces the initial conditions of heavy ion collision.
In fact, strangeness neutrality would rather impose $\mu_s \alt \mu_l$.
In general, the setup $\mu_l=\mu_s= \mu_B/3$ approximates strangeness neutrality at low temperatures, while
the $\mu_s= 0$ setup is relevant for high enough temperatures. \\
 It is natural  to expect that the effect of taking $\mu_s=\mu_l$ instead of $\mu_s=0$
becomes less and less evident when $\mu_l/T$ approaches zero, so that the 
curvature $\kappa$ at zero baryon density should not differ too much in the
two cases. The numerical analysis of Refs.~\cite{ Bonati:2014rfa,Bonati:2015bha,Bellwied:2015rza} has shown 
that this effect is invisible within the accuracy of the lattice setup
adopted there.

The paper is organized as follows: in Sec.~II we give some further details
of our numerical simulations. In Sec.~III we show our numerical results
for $\kappa$. Finally, in Sec.IV we draw our conclusions.

\section{Simulation details and numerical results}

We perform simulations of lattice QCD with 2+1 flavors of rooted staggered 
quarks at imaginary quark chemical potential.
We have made use of the HISQ/tree action~\cite{Follana:2006rc,Bazavov:2009bb,Bazavov:2010ru} 
as implemented in the publicly available MILC code~\cite{MILC}, 
which has been suitably modified by us in order to introduce an imaginary quark
chemical potential $\mu = \mu_B/3$. 
That has been done by multiplying all forward and backward 
temporal links entering the discretized Dirac operator 
by $\exp (i a \mu )$ and $\exp (- i a \mu )$, respectively: in this way,
the fermion determinant is still real and positive, so that standard
Monte Carlo methods can be applied.
As already remarked above, in the present study we have 
$\mu=\mu_l=\mu_s$. This means that the Euclidean partition function of the
discretized theory reads
\begin{equation}
\label{Z}
Z=\int [DU] e^{-S_{\rm gauge}} \prod_{q=u,d,s} {\rm det}(D_q[U,\mu])^{1/4} \;,
\end{equation}
where $S_{\rm gauge}$ is the Symanzik-improved gauge 
action and $D_q[U,\mu]$ is the staggered Dirac operator, modified as 
explained above for the inclusion of the imaginary quark chemical
(see Ref.~\cite{Bazavov:2009bb} and appendix~A of Ref.~\cite{Bazavov:2010ru}    for the precise definition of
the gauge action and the covariant derivative for highly improved staggered fermions). 

All simulations make use of the rational hybrid Monte Carlo (RHMC) 
algorithm. The length of each RHMC trajectory has been set to  
$1.0$ in molecular dynamics time units.

We have simulated the theory at finite temperature, and for several values of 
the imaginary quark chemical potential, near the transition temperature, 
adopting lattices of size $16^3\times 6$, $24^3\times 6$, $32^3 \times 8$, 
$40^3 \times 10$ and $48^3 \times 12$.  
We have discarded typically not less than one thousand trajectories for each 
run and have collected from {4k to 8k} trajectories for measurements.

\begin{figure}[tb]
\includegraphics*[width=1.\columnwidth]
{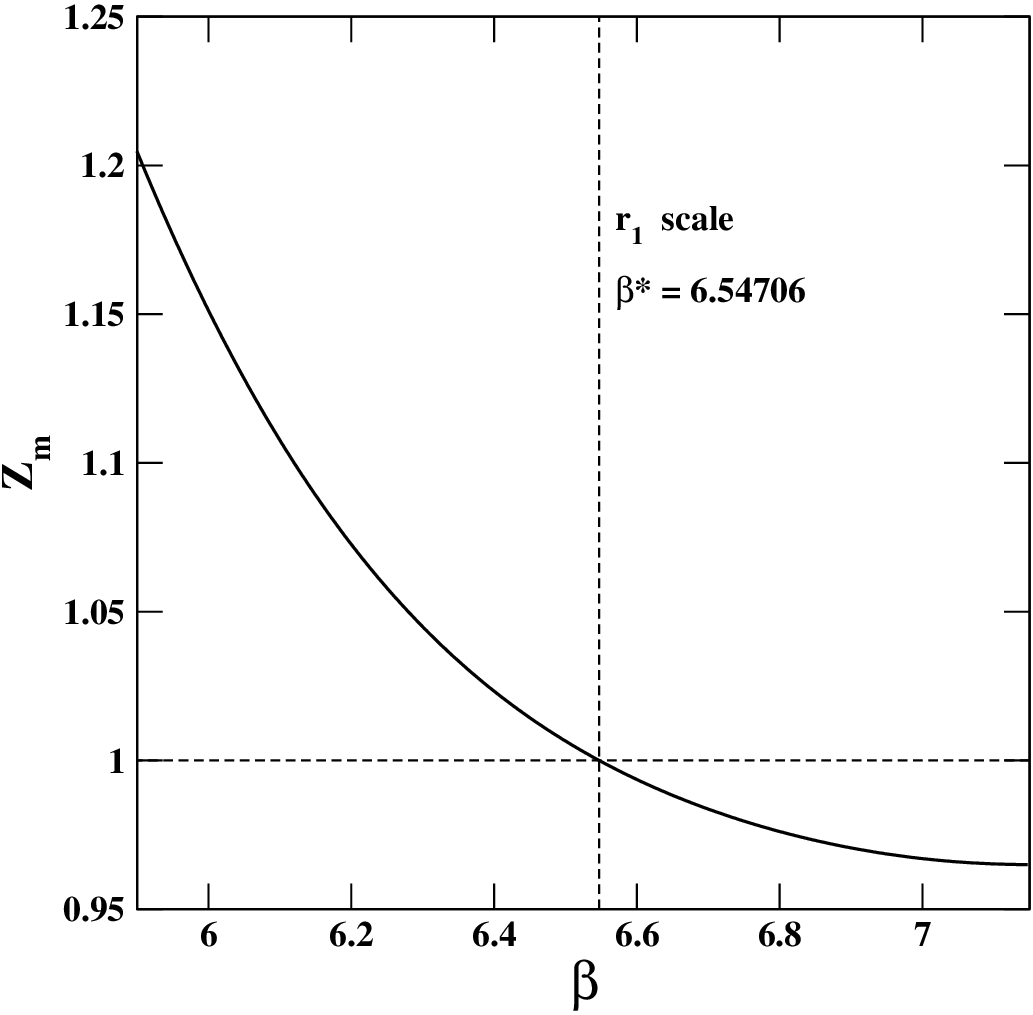}
\caption{The multiplicative renormalization factor $Z_m$ in the case of 
$r_1$-scale. The renormalization point is $\beta^*=6.54706$.}
\label{fig_zetam}
\end{figure}

\begin{figure}[tb]
\includegraphics*[width=1.\columnwidth]
{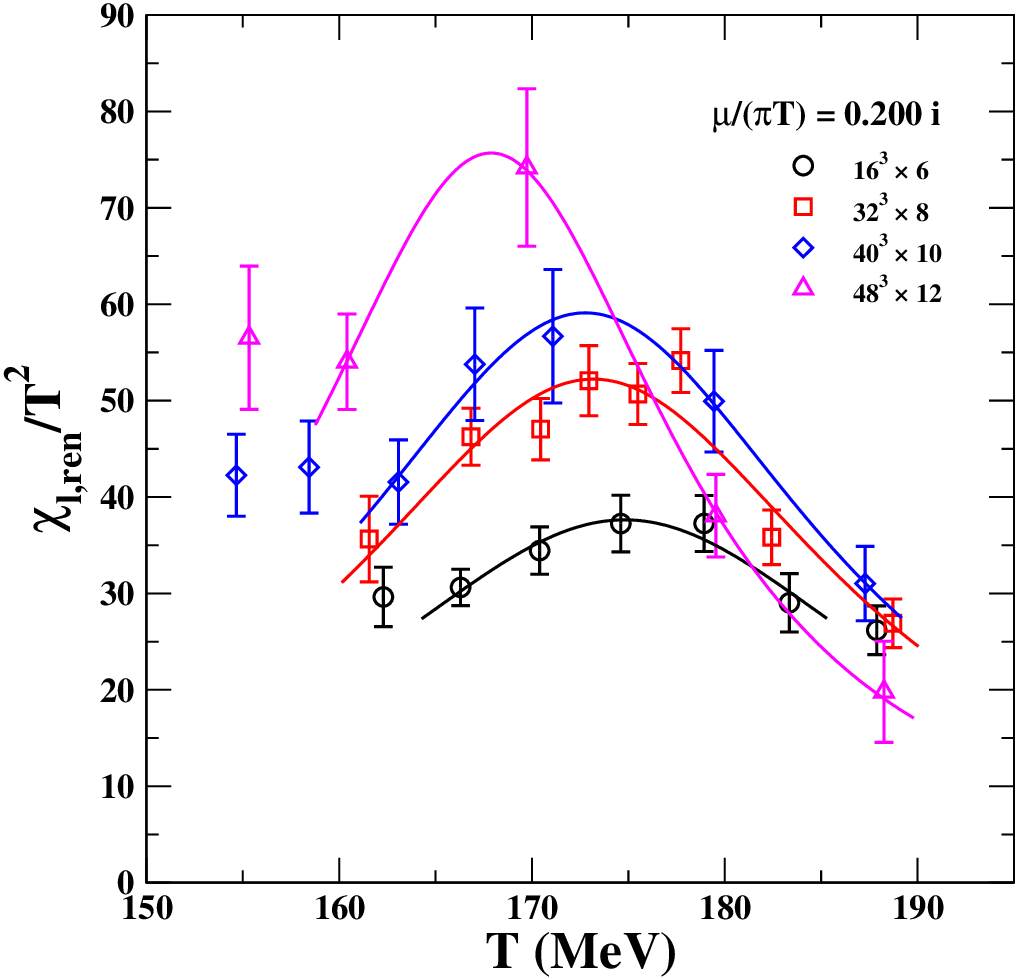}
\caption{The real part of the renormalized susceptibility of the light quark 
chiral condensate over $T^2$ on the lattices $16^3\times 6$,
$32^3\times 8$, $40^3\times 10$ and $48^3\times 12$ at $\mu/(\pi T)=0.2i$. 
Full lines give the Lorentzian fits near the peaks. The temperature
has been determined from the $r_1$ scale.}
\label{fig_chiral_light_suscep}
\end{figure}

The pseudocritical point $\beta_c(\mu^2)$ has been determined as the value 
for which the renormalized disconnected susceptibility of the light quark chiral condensate
divided by $T^2$ exhibits a peak. \\
The bare disconnected susceptibility is given by:
\begin{equation}
\label{chi_dis}
\chi_{l, \rm disc} =
{{n_f^2} \over 16 L_s^3 L_t}\left\{
\langle\bigl( {\rm Tr} D_q^{-1}\bigr)^2  \rangle -
\langle {\rm Tr} D_q^{-1}\rangle^2 \right\}\;,
\end{equation}
 Here $n_f=2$ is the number of light flavors
and $L_s$ denotes the lattice size in the space direction.  The renormalized chiral susceptibility
is defined as: 
\begin{equation}
\label{chi_ren}
\chi_{l, \rm ren} =   \frac{1}{Z_m^2} \;  \chi_{l, \rm disc} .
\end{equation}
The multiplicative renormalization factor $Z_m$  can be deduced
from an analysis of the line of constant physics for the light quark masses.
More precisely, we have~\cite{Bazavov:2010ru}:
\begin{equation}
\label{zeta}
Z_m(\beta) \;  =  \;  \frac{m_l(\beta)}{m_l(\beta^*)}  \; ,
\end{equation}
where the  renormalization point $\beta^*$ is chosen such that:
\begin{equation}
\label{beta*}
\frac{r_1}{a(\beta^*)} \; =  \;  2.37 \; ,
\end{equation}
where the function  $a(\beta)$ is discussed below.
In Fig.~\ref{fig_zetam} is shown the multiplicative renormalization 
factor $Z_m$ determined in the case  when $r_1$ is used to set the
scale (see below).
To precisely  localize the peak in $\chi_{l, \rm ren}/T^2$, a Lorentzian fit has 
been used. For illustrative purposes, 
in Fig.~\ref{fig_chiral_light_suscep} we display our determination of the 
pseudocritical couplings at $\mu/(\pi T)=0.2i$ for all lattices
considered in this work. The complete collection of results for
the disconnected susceptibility of the light quark chiral condensate obtained
in this work is presented in the Appendix.

To get the ratios $T_c(\mu)/T_c(0)$, we fix the lattice spacing through the 
observables $r_1$ and $f_K$, following the discussion in the Appendix~B of 
Ref.~\cite{Bazavov:2011nk}. \\
 For the  $r_1$ scale  the lattice spacing
is given in terms of the $r_1$ parameter as:
\begin{equation}
\label{scale-r1}
\frac{a}{r_1}(\beta)_{m_l=0.05m_s}=
\frac{c_0 f(\beta)+c_2 (10/\beta) f^3(\beta)}{
1+d_2 (10/\beta) f^2(\beta)} \; ,
\end{equation}
with $c_0=44.06$, $c_2=272102$, $d_2=4281$, $r_1=0.3106(20)\ {\text{fm}}$. \\
On the other hand, in the case of  the  $f_K$ scale we have:
 \begin{equation}
\label{scale-fk}
 a f_K(\beta)_{m_l=0.05m_s}=
\frac{c_0^K f(\beta)+c_2^K (10/\beta) f^3(\beta)}{
1+d_2^K (10/\beta) f^2(\beta)} \; ,
\end{equation}
with $c_0^K=7.66$, $c_2^K=32911$, $d_2^K=2388$, $r_1f_K \simeq 0.1738$. 
In Eqs.~(\ref{scale-r1}) and (\ref{scale-fk}), $f(\beta)$ is the two-loop beta 
function,
\begin{equation}
\label{beta function}
f(\beta)=(b_0 (10/\beta))^{-b_1/(2 b_0^2)} \exp(-\beta/(20 b_0))\;,
\end{equation}
$b_0$ and $b_1$ being its universal coefficients. \\

Our results are summarized in Table~\ref{summary}.
For all lattice sizes but $24^3\times6$ (where we have only one value of $\mu$),
the behavior of $T_c(\mu)/T_c(0)$ can be nicely fitted with a linear function 
in $\mu^2$,
\begin{equation}
\label{linearfit}
\frac{T_c(\mu)}{T_c(0)} = 1 + R_q \left(\frac{i \mu}{\pi T_c(\mu)}\right)^2 \;,
\end{equation}
which gives us access to the curvature $R_q$ and, hence, to the curvature
parameter $\kappa=-R_q/(9\pi^2)$ introduced in Eq.~(\ref{curv}). 
On the $24^3\times6$ lattice the linearity in $\mu^2$ has been assumed to hold, 
in order to extract $R_q$ from the only available determination at 
$\mu/(\pi T)=0.2i$.

\begin{table}[tb]
\setlength{\tabcolsep}{0.9pc}
\centering
\caption[]{Summary of the values of the ratio $T_c(\mu)/T_c(0)$ for 
the imaginary quark chemical potentials $\mu$ considered in this work.
The data for $\mu=0$ on the $24^3\times6$, $32^3\times68$ and $48^3\times12$ 
lattices have been estimated from the disconnected chiral susceptibilities 
reported respectively on Tables~X, XI and XII of Ref.~\cite{Bazavov:2011nk}.
The datum for $\mu=0$ on the $40^3\times10$ lattice has been estimated from 
the disconnected chiral susceptibilities reported on Table~XI of 
Ref.~\cite{Bazavov:2014pvz}.
The values of $T_c(\mu)/T_c(0)$ evaluated fixing the lattice scale by
$r_1$ and $f_K$ are reported, respectively, in the third and in the 
fourth column of the table.}
\begin{tabular}{cllll}
\hline
\hline
lattice & $\mu/(\pi T)$ & $T_c(\mu)/T_c(0)$  & $T_c(\mu)/T_c(0)$   \\
        &               & \ ($r_1$ scale)      & \ ($f_K$ scale) \\
\hline
16$^3\times 6$ & 0.15$i$ & 1.038(13) & 1.043(14) \\
               & 0.2$i$  & 1.063(15) & 1.070(15) \\
               & 0.25$i$ & 1.085(16) & 1.095(18) \\
\hline
24$^3\times 6$ & 0.2$i$  & 1.061(9)  & 1.067(10) \\
\hline
32$^3\times 8$ & 0.15$i$ & 1.054(7)  & 1.059(8) \\
               & 0.2$i$  & 1.066(10) & 1.071(11) \\
               & 0.25$i$ & 1.117(10) & 1.126(10) \\
\hline
40$^3\times 10$ & 0.15$i$ & 1.023(23)  & 1.024(24) \\
                & 0.2$i$  & 1.075(14)  & 1.079(15) \\
                & 0.25$i$ & 1.102(15)  & 1.107(15) \\
\hline
48$^3\times 12$ & 0.15$i$  & 1.013(31)  & 1.013(33) \\
                & 0.20$i$  & 1.051(14)  & 1.052(15) \\
                & 0.25$i$  & 1.094(26)  & 1.097(25) \\
\hline
\hline
\end{tabular}
\label{summary}
\end{table}

\begin{figure}[tb]
\includegraphics*[width=1.\columnwidth]
{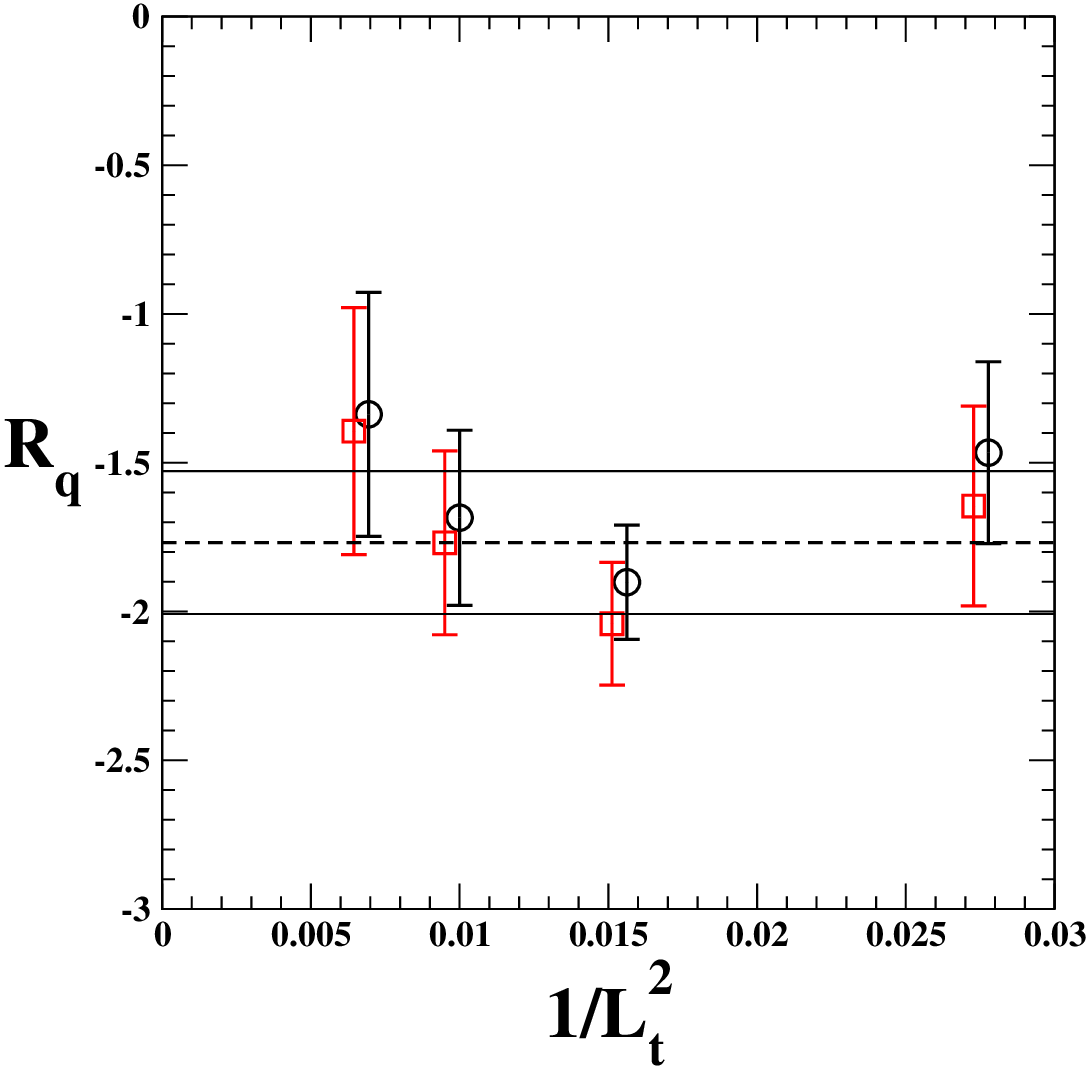}
\caption{Determinations of the curvature $R_q$ on the lattices $24^3\times 6$, 
$32^3 \times 8$, $40^3 \times 10$, $48^3 \times 12$, and from the two different 
methods to set the scale, {\it versus} $1/L_t^2$. Data points related with 
the $f_K$ scale setting have been slightly shifted along the horizontal axis 
for better readability. The dashed horizontal line gives the result of 
the fit to all data with a constant; the solid horizontal lines indicate the 
uncertainty on this constant (95\% confidence level).}

\label{curvature_vs_Nt}
\end{figure}

For the sake of the extrapolation to the continuum limit, in 
Fig.~\ref{curvature_vs_Nt} we report our determinations of $R_q$ on the 
lattices $24^3\times 6$, $32^3\times 8$, $40^3\times 10$, $48^3\times 12$,  
and from the two different methods to set the scale, {\it versus} $1/L_t^2$.

Within our accuracy, cutoff effects on $R_q$ are negligible, so that a 
constant fit works well over the whole region ($\chi^2_r \simeq 0.99$), thus 
including also the smallest $24^3\times6$ lattice. Taking into account
the uncertainties due to the continuum limit extrapolation, 
\begin{equation}
\label{curvature}
\kappa = 0.020(4) \;.
\end{equation}
Our estimate of the uncertainties for the curvature given in Eq.~(\ref{curvature}) takes into account both the error in the fit minimization
and the choice of the minimization function.
We stress, however, that if we exclude from the fit the value on the lattice
with the smallest $L_t$, {\it i.e.} the rightmost points in 
Fig.~\ref{curvature_vs_Nt}, the extrapolation to the continuum becomes largely
undetermined. Indeed, with the values of $R_q$ obtained in the present work
(see Table~\ref{Rq}), the fit with a constant is rather stable even if
$L_t=6$ is excluded, but the fit with a linear function in $1/L_t^2$ in the 
latter case gives a much smaller value of the curvature $\kappa$, though
with a large uncertainty (see Table~\ref{fit}).
\begin{table*}[tb]
\setlength{\tabcolsep}{0.9pc}
\centering
\caption[]{Summary of determinations of the curvature $R_q$ for all values of 
$L_t$ considered in this work, and from the two different methods to set the 
scale.}
\begin{tabular}{lcccc}
\hline
\hline
$L_t$ & $6$ & $8$  & $10$  & $12$\\
\hline
$R_q$ ($r_1$ scale) & $-1.466(306)$ & $-1.902(192)$ & $-1.685(294)$ 
& $-1.337(410)$ \\
$\kappa$ ($r_1$ scale) & $0.017(3)$ &  $0.021(2) $  & $0.019(3)$ & $0.015(5)$ \\
\hline
$R_q$ ($f_K$ scale) & $-1.646(336)$ & $-2.041(206)$ & $-1.769(309)$ 
& $-1.394(415)$ \\
$\kappa$ ($r_K$ scale) & $0.019(4)$ & $0.023(2)$ & $0.020(3)$ & $0.016(5)$ \\
\hline
\hline
\end{tabular}
\label{Rq}
\end{table*}
\begin{table*}[tb]
\setlength{\tabcolsep}{1.5pc}
\centering
\caption[]{Summary of the fit of the curvature $\kappa$ with the 
function $\kappa(L_t) = \kappa + A/L_t^2$, with $A$ taken equal to zero or left
free. The first column specifies the values of $L_t$ included in the fit,
the last column the reduced $\chi^2$. The uncertainties on the fit parameters
are obtained with 70\% confidence level.}
\begin{tabular}{lllc}
\hline
\hline
$L_t$ included & \hspace{0.6cm} $\kappa$ & \hspace{0.6cm} $A$ & $\chi^2_{\rm r}$ 
\\
\hline
6, 8, 10, 12 & 0.01991(114) & 0.            & 0.76 \\ 
6, 8, 10, 12 & 0.02014(449) & \hspace{-0.35cm} $-$0.015(259)  & 0.89 \\ 
\hline
8, 10, 12 & 0.02048(127) & 0.         & 0.80 \\ 
8, 10, 12 & 0.01182(748) & 0.669(559) & 0.14 \\ 
\hline
\hline
\end{tabular}
\label{fit}
\end{table*}

\section{Conclusions and discussion}

We have studied QCD with $n_f = 2+1$ flavors discretized in the HISQ/tree 
rooted staggered fermion formulation and in the presence of an imaginary 
baryon chemical potential, with a physical strange quark mass and a 
light-to-strange mass ratio $m_l/m_s = 1/20$, and $\mu=\mu_l=\mu_s$.

We have estimated, by the method of analytic continuation, the continuum limit
of the curvature of the pseudocritical line in the temperature - baryon 
chemical potential, defined in Eq.~(\ref{curv}). The observable adopted
to identify, for each fixed $\mu$, the crossover temperature has been
the disconnected part of the renormalized susceptibility of the light quark 
chiral condensate, in units of the squared temperature. This observable is
convenient for many reasons: it dominates, for small enough quark masses, the
whole light chiral susceptibility, which would be much harder to implement;
it undergoes only a multiplicative renormalization; it is strongly sensitive 
to the transition, thus allowing precise determinations of the pseudocritical 
temperatures.

We have found that, within the accuracy of our determinations, cutoff effects
on the curvature are negligible already on the lattice with temporal size
$L_t=6$. Our determination of the curvature parameter, $\kappa$=0.020(4),
is indeed compatible with the value quoted in our previous 
paper~\cite{Cea:2014xva}, $\kappa$=0.018(4), without the extrapolation to
the continuum.

\begin{figure}[tb]
\includegraphics*[height=0.295\textheight,width=1.\columnwidth]
{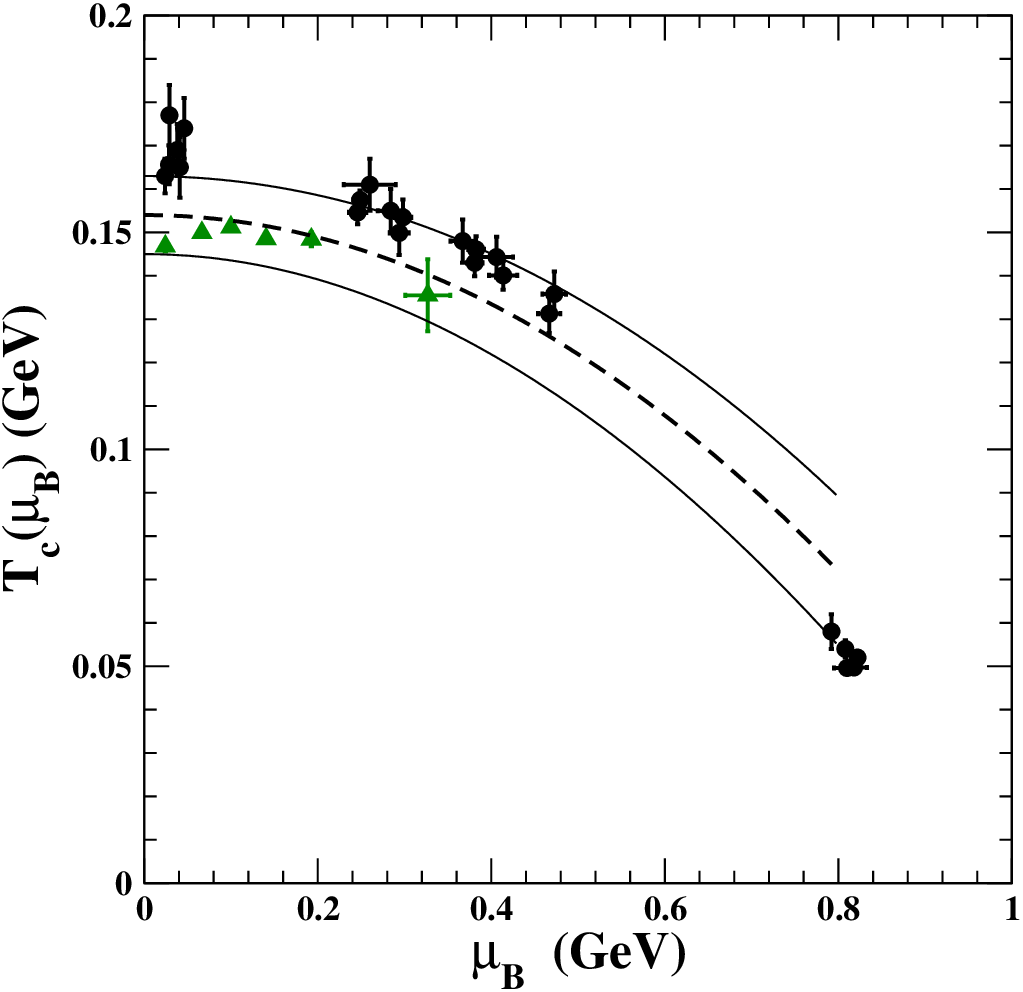}
\caption{$T_c(\mu_B)$ versus $\mu_B$ (units in GeV). 
Experimental values of $T_c(\mu_B)$ are taken
from Fig.~1 of Ref.~\cite{Cleymans:2005xv} (black circles) and from
Fig.~3 of Ref.~\cite{Alba:2014eba} (green triangles),  for the 
standard  hadronization model and for the  susceptibilities of conserved charges respectively. 
The dashed line is a parametrization  corresponding to
$T_c(\mu_B) = T_c(0) - b \mu_B^2$ with $T_c(0)=0.154(9)\, \text{GeV}$ and 
$b=0.128(25)\,{\text{GeV}}^{-1}$ . 
The solid lines represent the corresponding  error band.}
\label{Tcmu}
\end{figure}

It is interesting to extrapolate the critical line as determined in this work 
to the region of real baryon density and compare it with the freeze-out
curves resulting from a few phenomenological analyses of relativistic
heavy-ion collisions. This is done in Fig.~\ref{Tcmu}, where we report
two different estimates. The first is from the analysis of 
Ref.~\cite{Cleymans:2005xv}, based on the standard statistical hadronization 
model, where the freeze-out curve is parametrized as 
\begin{equation}
\label{cleymans}
T_c(\mu_B) = a - b \mu_B^2 - c \mu_B^4\;,
\end{equation}
with $a=0.166(2) \ {\text{GeV}}$, $b=0.139(16) \ {\text{GeV}}^{-1}$, 
and $c=0.053(21) \ {\text{GeV}}^{-3}$.
The second estimate is from Ref.~\cite{Alba:2014eba} and is based on the analysis
of susceptibilities of the (conserved) baryon and electric charges.
In fact, our critical line is in nice agreement with all the freeze-out points of
Refs.~\cite{Cleymans:2005xv,Alba:2014eba}. In particular, using our estimate
of the curvature,  Eq.~(\ref{curvature}), we get   
$b=0.128(25) \ {\text{GeV}}^{-1}$, in very good agreement with the
quoted phenomenological  value. The significance of the comparison
presented in Fig.~\ref{Tcmu}, with special reference to the question
whether the pseudocritical line lies indeed above the freeze-out curve, can be 
increased at the (nonnegligible) price of reducing the uncertainties on 
$T_c(0)$ and on $\kappa$. \\

Some {\it caveats} are in order here. We do not expect our critical line 
to be reliable too far from $\mu=0$: as a rule of thumb, we can trust it up 
to real quark chemical potentials of the same order of the modulus
of the largest imaginary chemical potential included in the 
fit~(\ref{linearfit}), {\it i.e.} $|\mu|/(\pi T)=0.25$. This translates
to real baryon chemical potentials in the region $\mu_B \lesssim 0.4\,\text{GeV}$.
Moreover, the effect of taking $\mu_s=\mu_l$ instead of $\mu_s < \mu_l$
should become visible on the shape of the critical line as we move away
from $\mu=0$ in the region of real baryon densities, thus reducing
further the region of reliability of our critical line. So, from a prudential
point of view, the agreement shown in Fig.~\ref{Tcmu} could be considered
the fortunate combination of different kinds of systematic effects. We cannot
however exclude the possibility that the message from Fig.~\ref{Tcmu} is to be 
interpreted in positive sense, {\it i.e.} the setup we adopted and the
observable we considered may catch better some features of the crossover
transition, thus explaining the nice comparison with freeze-out data.
Indeed,  our  result for the  continuum extrapolation of the curvature $\kappa$ is
in fair agreements with the recent estimates in  Ref.~\cite{Bonati:2015bha}, where
both setup  $\mu_s=\mu_l$ and $\mu_s=0$ were adopted,
and Ref.~\cite{ Bellwied:2015rza},  where the strangeness neutral trajectories
were determined from lattice simulations  by imposing $\langle n_S \rangle=0$.

\section*{Acknowledgments}
This work was in part based on the MILC collaboration's public lattice gauge 
theory code. See {\url{http://physics.utah.edu/~detar/milc.html}}.
This work has been partially supported by the INFN SUMA project.
Simulations have been performed on BlueGene/Q at CINECA 
(Projects Iscra-B/EXQCD and INF14\_npqcd), on the BC$^2$S cluster in Bari and on the CSNIV 
Zefiro cluster in Pisa.

\appendix*

\section{Appendix: Summary of data for the disconnected susceptibility of the 
light quark chiral condensate}

In this Appendix (Tables~\ref{data_16x6_0.150} through~\ref{data_48x12_0.250}), 
we summarize all the results for the disconnected
chiral susceptibility obtained in our simulations for each considered
value of the coupling $\beta$ and for the corresponding physical temperature,
as determined from the two different procedures to set the scale.

\begin{table*}[htb]
\setlength{\tabcolsep}{0.9pc}
\centering
\caption[]{Data for the disconnected chiral susceptibility
on the $16^3\times 6$ lattice at $\mu/(\pi T)=0.150i$: the second
column gives $\chi_{l,{\rm disc}}$ defined in Eq.~(\ref{chi_dis}), the
fourth and sixth columns give $\chi_{l,{\rm ren}}/T^2$, with $\chi_{l,{\rm ren}}$
defined in Eq.~(\ref{chi_ren}) and the temperature $T$ (columns three and 
five, respectively) determined by the two different methods to set the scale.}
\begin{tabular}{cccccc}
\hline
\hline
$\beta$ & $\chi_{l,{\text{disc}}}$  & \makecell{$T(\rm{MeV})$ \\ ($r_1$ scale)}
& $\chi_{l,{\text{ren}}}/T^2$ & \makecell{$T(\rm{MeV})$ \\ ($f_K$ scale)}
& $\chi_{l,{\text{ren}}}/T^2$ \\
\hline
 6.000  &  0.939 (0.114)  &  147.3  &  25.5 (3.1) &  138.2  &  27.7 (3.4)  \\
 6.050  &  0.495 (0.134)  &  154.6  &  14.0 (3.8) &  145.8  &  15.0 (4.1)  \\
 6.100  &  1.133 (0.145)  &  162.3  &  33.2 (4.3) &  153.8  &  35.4 (4.5)  \\
 6.125  &  1.260 (0.089)  &  166.3  &  37.6 (2.6) &  158.0  &  39.8 (2.8)  \\
 6.150  &  1.273 (0.079)  &  170.4  &  38.6 (2.4) &  162.3  &  40.7 (2.5)  \\
 6.175  &  1.187 (0.086)  &  174.6  &  36.6 (2.7) &  166.7  &  38.4 (2.8)  \\
 6.195  &  1.093 (0.107)  &  178.1  &  34.1 (3.3) &  170.2  &  35.6 (3.5)  \\
 6.200  &  0.944 (0.100)  &  178.9  &  29.5 (3.1) &  171.1  &  30.8 (3.3)  \\
\hline
\hline
\end{tabular}
\label{data_16x6_0.150}
\end{table*}

\begin{table*}[htb]
\setlength{\tabcolsep}{0.9pc}
\centering
\caption[]{Data for the disconnected chiral susceptibility 
on the $16^3\times 6$ lattice at $\mu/(\pi T)=0.200i$ (legend as in 
Table~\ref{data_16x6_0.150}).}
\begin{tabular}{cccccc}
\hline
\hline
$\beta$ & $\chi_{l,{\text{disc}}}$  & \makecell{$T(\rm{MeV})$ \\ ($r_1$ scale)}
& $\chi_{l,{\text{ren}}}/T^2$ & \makecell{$T(\rm{MeV})$ \\ ($f_K$ scale)}
& $\chi_{l,{\text{ren}}}/T^2$ \\
\hline
 6.100  &  1.010 (0.105)  &  162.3  &  29.6 (3.1)  &  153.8  &  31.5 (3.3)  \\ 
 6.125  &  1.026 (0.063)  &  166.3  &  30.6 (1.9)  &  158.0  &  32.4 (2.0)  \\ 
 6.150  &  1.135 (0.081)  &  170.4  &  34.4 (2.5)  &  162.3  &  36.3 (2.6)  \\ 
 6.175  &  1.208 (0.095)  &  174.6  &  37.2 (2.9)  &  166.7  &  39.1 (3.1)  \\ 
 6.200  &  1.191 (0.093)  &  178.9  &  37.3 (2.9)  &  171.1  &  38.9 (3.0)  \\ 
 6.225  &  0.915 (0.095)  &  183.4  &  29.0 (3.0)  &  175.8  &  30.2 (3.2)  \\ 
 6.250  &  0.814 (0.078)  &  187.9  &  26.2 (2.5)  &  180.5  &  27.1 (2.6)  \\ 
\hline
\hline
\end{tabular}
\label{data_16x6_0.200}
\end{table*}

\begin{table*}[htb]
\setlength{\tabcolsep}{0.9pc}
\centering
\caption[]{Data for the disconnected chiral susceptibility 
on the $16^3\times 6$ lattice at $\mu/(\pi T)=0.250i$ (legend as in 
Table~\ref{data_16x6_0.150}).}
\begin{tabular}{cccccc}
\hline
\hline
$\beta$ & $\chi_{l,{\text{disc}}}$  & \makecell{$T(\rm{MeV})$ \\ ($r_1$  scale)}   
& $\chi_{l,{\text{ren}}}/T^2$ & $T(\rm{MeV})$ ($f_K$  scale)   
& $\chi_{l,{\text{ren}}}/T^2$ \\
\hline
 6.150  &  0.899 (0.095)  &  170.4  &  27.3 (2.9)  &  162.3  &  28.8 (3.0)  \\ 
 6.180  &  1.089 (0.064)  &  175.5  &  33.7 (2.0)  &  167.5  &  35.3 (2.1)  \\ 
 6.186  &  1.078 (0.083)  &  176.5  &  33.5 (2.6)  &  168.6  &  35.0 (2.7)  \\  
 6.200  &  1.117 (0.097)  &  178.9  &  34.9 (3.0)  &  171.1  &  36.5 (3.2)  \\ 
 6.250  &  0.784 (0.066)  &  187.9  &  25.2 (2.1)  &  180.5  &  26.1 (2.2)  \\ 
 6.300  &  0.561 (0.051)  &  197.3  &  18.5 (1.7)  &  190.2  &  19.0 (1.7)  \\ 
 6.350  &  0.189 (0.047)  &  207.2  &   6.4 (1.6)  &  200.5  &   6.5 (1.6)  \\ 
\hline
\hline
\end{tabular}
\label{data_16x6_0.250}
\end{table*}

\begin{table*}[htb]
\setlength{\tabcolsep}{0.9pc}
\centering
\caption[]{Data for the disconnected chiral susceptibility 
on the $24^3\times 6$ lattice at $\mu/(\pi T)=0.200i$ (legend as in 
Table~\ref{data_16x6_0.150}).}
\begin{tabular}{cccccc}
\hline
\hline
$\beta$ & $\chi_{l,{\text{disc}}}$  & \makecell{$T(\rm{MeV})$ \\ ($r_1$  scale)}   
& $\chi_{l,{\text{ren}}}/T^2$ & \makecell{$T(\rm{MeV})$ \\ ($f_K$ scale)}   
& $\chi_{l,{\text{ren}}}/T^2$ \\
\hline
 6.120  &  0.916 (0.107)  &  165.5  &  27.3 (3.2)  & 157.2  &  28.9 (3.4) \\
 6.150  &  1.106 (0.136)  &  170.4  &  33.6 (4.1)  & 162.3  &  35.4 (4.4) \\
 6.165  &  0.849 (0.079)  &  172.9  &  26.0 (2.4)  & 164.9  &  27.3 (2.5) \\
 6.180  &  1.084 (0.073)  &  175.5  &  33.5 (2.3)  & 167.5  &  35.1 (2.4) \\
 6.195  &  1.288 (0.078)  &  178.1  &  40.2 (2.4)  & 170.2  &  42.0 (2.5) \\
 6.210  &  1.159 (0.087)  &  180.7  &  36.5 (2.7)  & 173.0  &  38.0 (2.9) \\
 6.225  &  1.244 (0.098)  &  183.4  &  39.5 (3.1)  & 175.8  &  41.1 (3.2) \\
 6.240  &  1.030 (0.070)  &  186.1  &  33.0 (2.2)  & 178.6  &  34.2 (2.3) \\
 6.255  &  0.979 (0.160)  &  188.8  &  31.6 (5.1)  & 181.4  &  32.7 (5.3) \\
 6.270  &  0.692 (0.062)  &  191.6  &  22.5 (2.0)  & 184.3  &  23.2 (2.1) \\
 6.300  &  0.430 (0.038)  &  197.3  &  14.2 (1.2)  & 190.2  &  14.6 (1.3) \\
\hline
\hline
\end{tabular}
\label{data_24x6_0.200}
\end{table*}

\begin{table*}[htb]
\setlength{\tabcolsep}{0.9pc}
\centering
\caption[]{Data for the disconnected chiral susceptibility 
on the $32^3\times 8$ lattice at $\mu/(\pi T)=0.150i$ (legend as in 
Table~\ref{data_16x6_0.150}).}
\begin{tabular}{cccccc}
\hline
\hline
$\beta$ & $\chi_{l,{\text{disc}}}$  & \makecell{$T(\rm{MeV})$ \\ ($r_1$  scale)}   
& $\chi_{l,{\text{ren}}}/T^2$ & \makecell{$T(\rm{MeV})$ \\ ($f_K$ scale)}   
& $\chi_{l,{\text{ren}}}/T^2$ \\
\hline
 6.415  & 0.646 (0.073)  &  165.5  &  39.7 (4.5)  &  160.9  &  40.1 (4.6) \\ 
 6.424  & 0.730 (0.047)  &  167.0  &  45.0 (2.9)  &  162.4  &  45.5 (3.0) \\ 
 6.429  & 0.780 (0.035)  &  167.8  &  48.2 (2.2)  &  163.2  &  48.6 (2.2) \\ 
 6.436  & 0.865 (0.059)  &  169.0  &  53.6 (3.7)  &  164.4  &  54.1 (3.7) \\ 
 6.450  & 0.962 (0.047)  &  171.3  &  59.8 (3.0)  &  166.8  &  60.3 (3.0) \\ 
 6.460  & 0.883 (0.070)  &  173.0  &  55.1 (4.4)  &  168.6  &  55.4 (4.4) \\ 
 6.470  & 0.776 (0.060)  &  174.6  &  48.6 (3.8)  &  170.3  &  48.8 (3.8) \\ 
\hline
\hline
\end{tabular}
\label{data_32x8_0.150}
\end{table*}

\begin{table*}[htb]
\setlength{\tabcolsep}{0.9pc}
\centering
\caption[]{Data for the disconnected chiral susceptibility 
on the $32^3\times 8$ lattice at $\mu/(\pi T)=0.200i$ (legend as in 
Table~\ref{data_16x6_0.150}).}
\begin{tabular}{cccccc}
\hline
\hline
$\beta$ & $\chi_{l,{\text{disc}}}$  & \makecell{$T(\rm{MeV})$ \\ ($r_1$  scale)}   
& $\chi_{l,{\text{ren}}}/T^2$ & \makecell{$T(\rm{MeV})$ \\ ($f_K$ scale)}   
& $\chi_{l,{\text{ren}}}/T^2$ \\
\hline
 6.390  &  0.585 (0.073)  &  161.7  &  35.6 (4.4)  &  156.8  &  36.2 (4.5) \\
 6.423  &  0.751 (0.048)  &  166.8  &  46.3 (3.0)  &  162.2  &  46.8 (3.0) \\
 6.445  &  0.757 (0.051)  &  170.4  &  47.0 (3.2)  &  166.0  &  47.4 (3.2) \\
 6.460  &  0.834 (0.058)  &  173.0  &  52.1 (3.6)  &  168.6  &  52.4 (3.7) \\
 6.475  &  0.808 (0.051)  &  175.5  &  50.7 (3.2)  &  171.2  &  50.9 (3.2) \\
 6.488  &  0.860 (0.053)  &  177.7  &  54.2 (3.3)  &  173.5  &  54.3 (3.3) \\
 6.515  &  0.565 (0.045)  &  182.4  &  35.8 (2.8)  &  178.3  &  35.8 (2.8) \\
 6.550  &  0.420 (0.039)  &  188.7  &  26.9 (2.5)  &  184.8  &  26.8 (2.5) \\
\hline
\hline
\end{tabular}
\label{data_32x8_0.200}
\end{table*}

\begin{table*}[htb]
\setlength{\tabcolsep}{0.9pc}
\centering
\caption[]{Data for the disconnected chiral susceptibility 
on the $32^3\times 8$ lattice at $\mu/(\pi T)=0.250i$ (legend as in 
Table~\ref{data_16x6_0.150}).}
\begin{tabular}{cccccc}
\hline
\hline
$\beta$ & $\chi_{l,{\text{disc}}}$  & \makecell{$T(\rm{MeV})$ \\ ($r_1$  scale)}   
& $\chi_{l,{\text{ren}}}/T^2$ & \makecell{$T(\rm{MeV})$ \\ ($f_K$ scale)}   
& $\chi_{l,{\text{ren}}}/T^2$ \\
\hline
 6.420  &  0.613 (0.055)  &  166.4  &  37.7 (3.4)  &  161.7  &  38.2 (3.4) \\
 6.455  &  0.555 (0.044)  &  172.1  &  34.6 (2.7)  &  167.7  &  34.8 (2.7) \\
 6.490  &  0.792 (0.053)  &  178.1  &  49.9 (3.4)  &  173.8  &  50.0 (3.4) \\
 6.525  &  0.753 (0.051)  &  184.2  &  47.8 (3.3)  &  180.2  &  47.8 (3.3) \\
 6.560  &  0.568 (0.030)  &  190.6  &  36.5 (2.0)  &  186.7  &  36.3 (1.9) \\
\hline
\hline
\end{tabular}
\label{data_32x8_0.250}
\end{table*}

\begin{table*}[htb]
\setlength{\tabcolsep}{0.9pc}
\centering
\caption[]{Data for the disconnected chiral susceptibility 
on the $40^3\times 10$ lattice at $\mu/(\pi T)=0.150i$ (legend as in 
Table~\ref{data_16x6_0.150}).}
\begin{tabular}{cccccc}
\hline
\hline
$\beta$ & $\chi_{l,{\text{disc}}}$  & \makecell{$T(\rm{MeV})$ \\ ($r_1$  scale)}   
& $\chi_{l,{\text{ren}}}/T^2$ & \makecell{$T(\rm{MeV})$ \\ ($f_K$ scale)}   
& $\chi_{l,{\text{ren}}}/T^2$ \\
\hline
 6.550  &  0.444 (0.043)  &  151.0  &  44.4 (4.3)  &  147.9  &  44.3 (4.3) \\
 6.606  &  0.570 (0.070)  &  159.4  &  57.9 (7.1)  &  156.5  &  57.4 (7.0) \\
 6.648  &  0.524 (0.037)  &  165.9  &  53.6 (3.8)  &  163.2  &  53.0 (3.8) \\
 6.690  &  0.525 (0.048)  &  172.7  &  54.1 (4.9)  &  170.2  &  53.3 (4.8) \\
 6.732  &  0.402 (0.035)  &  179.8  &  41.7 (3.6)  &  177.4  &  41.0 (3.5) \\
\hline
\hline
\end{tabular}
\label{data_40x10_0.150}
\end{table*}

\begin{table*}[htb]
\setlength{\tabcolsep}{0.9pc}
\centering
\caption[]{Data for the disconnected chiral susceptibility 
on the $40^3\times 10$ lattice at $\mu/(\pi T)=0.200i$ (legend as in 
Table~\ref{data_16x6_0.150}).}
\begin{tabular}{cccccc}
\hline
\hline
$\beta$ & $\chi_{l,{\text{disc}}}$  & \makecell{$T(\rm{MeV})$ \\ ($r_1$  scale)}   
& $\chi_{l,{\text{ren}}}/T^2$ & \makecell{$T(\rm{MeV})$ \\ ($f_K$ scale)}   
& $\chi_{l,{\text{ren}}}/T^2$ \\
\hline
 6.575  &  0.420 (0.042)  &  154.7  &  42.3 (4.3)  &  151.7  &  42.0 (4.2) \\
 6.600  &  0.425 (0.047)  &  158.4  &  43.1 (4.8)  &  155.5  &  42.7 (4.7) \\
 6.630  &  0.407 (0.043)  &  163.1  &  41.6 (4.4)  &  160.3  &  41.1 (4.3) \\
 6.655  &  0.525 (0.057)  &  167.0  &  53.8 (5.8)  &  164.4  &  53.1 (5.8) \\
 6.680  &  0.550 (0.067)  &  171.1  &  56.7 (6.9)  &  168.5  &  55.8 (6.8) \\
 6.730  &  0.481 (0.051)  &  179.4  &  49.9 (5.3)  &  177.1  &  49.0 (5.2) \\
 6.775  &  0.297 (0.037)  &  187.3  &  31.0 (3.9)  &  185.1  &  30.4 (3.8) \\
\hline
\hline
\end{tabular}
\label{data_40x10_0.200}
\end{table*}

\begin{table*}[htb]
\setlength{\tabcolsep}{0.9pc}
\centering
\caption[]{Data for the disconnected chiral susceptibility 
on the $40^3\times 10$ lattice at $\mu/(\pi T)=0.250i$ (legend as in 
Table~\ref{data_16x6_0.150}).}
\begin{tabular}{cccccc}
\hline
\hline
$\beta$ & $\chi_{l,{\text{disc}}}$  & \makecell{$T(\rm{MeV})$ \\ ($r_1$  scale)}   
& $\chi_{l,{\text{ren}}}/T^2$ & \makecell{$T(\rm{MeV})$ \\ ($f_K$ scale)}   
& $\chi_{l,{\text{ren}}}/T^2$ \\
\hline
 6.645  &  0.386 (0.060)  &  165.4  &  39.5 (6.1)  & 162.7  &  39.0 (6.1) \\
 6.664  &  0.368 (0.041)  &  168.5  &  37.8 (4.2)  & 165.8  &  37.3 (4.1) \\
 6.714  &  0.552 (0.064)  &  176.7  &  57.2 (6.6)  & 174.3  &  56.3 (6.5) \\
 6.764  &  0.421 (0.059)  &  185.3  &  43.9 (6.1)  & 183.1  &  43.0 (6.0) \\
 6.810  &  0.193 (0.051)  &  193.6  &  20.2 (5.3)  & 191.5  &  19.8 (5.2) \\
\hline
\hline
\end{tabular}
\label{data_40x10_0.250}
\end{table*}

\begin{table*}[htb]
\setlength{\tabcolsep}{0.9pc}
\centering
\caption[]{Data for the disconnected chiral susceptibility 
on the $48^3\times 12$ lattice at $\mu/(\pi T)=0.150i$ (legend as in 
Table~\ref{data_16x6_0.150}).}
\begin{tabular}{cccccc}
\hline
\hline
$\beta$ & $\chi_{l,{\text{disc}}}$  & \makecell{$T(\rm{MeV})$ \\ ($r_1$  scale)}   
& $\chi_{l,{\text{ren}}}/T^2$ & \makecell{$T(\rm{MeV})$ \\ ($f_K$ scale)}   
& $\chi_{l,{\text{ren}}}/T^2$ \\
\hline
 6.700  &  0.242 (0.052) &  145.3  &  36.1 (7.7)  &  145.3  &  36.1 (7.7)  \\
 6.730  &  0.374 (0.047) &  149.5  &  55.9 (7.1)  &  149.5  &  55.9 (7.1)  \\
 6.780  &  0.433 (0.063) &  156.8  &  65.3 (9.5)  &  156.8  &  65.3 (9.5)  \\
 6.815  &  0.475 (0.111) &  162.1  &  71.9 (16.8) &  162.1  &  71.9 (16.8) \\
 6.830  &  0.382 (0.036) &  164.4  &  57.9 (5.5)  &  164.4  &  57.9 (5.5)  \\
 6.880  &  0.466 (0.068) &  172.3  &  71.1 (10.4) &  172.3  &  71.1 (10.4) \\
 6.930  &  0.247 (0.027) &  180.5  &  37.8 (4.1)  &  180.5  &  37.8 (4.1)  \\
 6.980  &  0.191 (0.032) &  189.1  &  29.3 (4.9)  &  189.1  &  29.3 (4.9)  \\
\hline
\hline
\end{tabular}
\label{data_48x12_0.150}
\end{table*}

\begin{table*}[htb]
\setlength{\tabcolsep}{0.9pc}
\centering
\caption[]{Data for the disconnected chiral susceptibility 
on the $48^3\times 12$ lattice at $\mu/(\pi T)=0.200i$ (legend as in 
Table~\ref{data_16x6_0.150}).}
\begin{tabular}{cccccc}
\hline
\hline
$\beta$ & $\chi_{l,{\text{disc}}}$  & \makecell{$T(\rm{MeV})$ \\ ($r_1$  scale)}   
& $\chi_{l,{\text{ren}}}/T^2$ & \makecell{$T(\rm{MeV})$ \\ ($f_K$ scale)}   
& $\chi_{l,{\text{ren}}}/T^2$ \\
\hline
 6.770  &  0.376 (0.049)  &  155.3  &  56.5 (7.4)  &  153.5  &  55.3 (7.3)  \\
 6.804  &  0.357 (0.033)  &  160.4  &  54.0 (5.0)  &  158.6  &  52.8 (4.8)  \\
 6.864  &  0.487 (0.054)  &  169.7  &  74.2 (8.2)  &  168.1  &  72.2 (8.0)  \\
 6.924  &  0.249 (0.028)  &  179.5  &  38.1 (4.3)  &  178.1  &  37.0 (4.2)  \\
 6.975  &  0.129 (0.034)  &  188.3  &  19.8 (5.2)  &  186.9  &  19.2 (5.1)  \\
\hline
\hline
\end{tabular}
\label{data_48x12_0.200}
\end{table*}

\begin{table*}[htb]
\setlength{\tabcolsep}{0.9pc}
\centering
\caption[]{Data for the disconnected chiral susceptibility 
on the $48^3\times 12$ lattice at $\mu/(\pi T)=0.250i$ (legend as in 
Table~\ref{data_16x6_0.150}).}
\begin{tabular}{cccccc}
\hline
\hline
$\beta$ & $\chi_{l,{\text{disc}}}$  & \makecell{$T(\rm{MeV})$ \\ ($r_1$  scale)}   
& $\chi_{l,{\text{ren}}}/T^2$ & \makecell{$T(\rm{MeV})$ \\ ($f_K$ scale)}
& $\chi_{l,{\text{ren}}}/T^2$ \\
\hline
 6.650  &  0.274 (0.044)  &  138.5  &  40.4 (6.5)  &  136.3  &  39.9 (6.4) \\
 6.750  &  0.175 (0.054)  &  152.4  &  26.3 (8.1)  &  150.5  &  25.8 (8.0) \\
 6.864  &  0.300 (0.042)  &  169.7  &  45.7 (6.3)  &  168.1  &  44.5 (6.2) \\
 6.900  &  0.418 (0.043)  &  175.6  &  64.0 (6.6)  &  174.1  &  62.2 (6.4) \\
 6.950  &  0.278 (0.037)  &  183.9  &  42.6 (5.6)  &  182.6  &  41.4 (5.4) \\
 6.975  &  0.224 (0.025)  &  188.3  &  34.4 (3.8)  &  186.9  &  33.4 (3.7) \\
 7.025  &  0.153 (0.018)  &  197.2  &  23.6 (2.7)  &  196.0  &  22.8 (2.6) \\
\hline
\hline
\end{tabular}
\label{data_48x12_0.250}
\end{table*}

\bibliography{hd_qcd}

\end{document}